\documentclass[conference,romanappendices,10pt]{IEEEtran}
\IEEEoverridecommandlockouts

\usepackage{algorithmic,amsmath,amssymb,amsthm,array,bbm,bibentry,cite,color,comment}
\usepackage{enumerate,enumitem,eurosym,float,graphicx,lettrine,mathrsfs,multirow,nomencl,import}
\usepackage{pict2e,psfrag,ragged2e,relsize,setspace,slashed,subfigure,supertabular,tabularx,url}
\usepackage[english]{babel}
\usepackage[T1]{fontenc}
\usepackage[utf8]{inputenc}

{		\newtheorem{theorem}{Theorem}}
{ 		\newtheorem{definition}{Definition}}
{ 		\newtheorem{lemma}{Lemma}}
{ 		}
{ 		\newtheorem{remark}{Remark}}
{		}
{		}
{ 		\newtheorem{corollary}{Corollary}}

\newcommand{\setF}{\mathcal{F}}

\newcommand{\setL}{\mathcal{L}}

\newcommand{\diff}{\mathrm{d}}
\newcommand{\Exp}{\mathbb{E}}

\renewcommand{\sc}{\textnormal{\tiny{SC}}}
\newcommand{\ul}{\textnormal{\tiny{UL}}}
\newcommand{\dl}{\textnormal{\tiny{DL}}}
\newcommand{\sir}{\mathsf{SIR}}
\newcommand{\Phit}{\mathsf{P}_{\mathrm{hit}}}
\newcommand{\Psuc}{\mathsf{P}_{\mathrm{suc}}}
\newcommand{\PsucLB}{\underline{\mathsf{P}}_{\mathrm{suc}}}

\title{Cache-Aided Full-Duplex Small Cells}
\author{\IEEEauthorblockN{Marco Maso,$^{1}$ Italo Atzeni,$^{1}$ Imène Ghamnia,$^{2}$ Ejder Ba\c{s}tu\u{g},$^{2,3}$ and Mérouane Debbah$^{1,2}$}
	\IEEEauthorblockA{${}^{1}$Mathematical and Algorithmic Sciences Lab, France Research Center, Huawei Technologies France SASU  \\
		${}^{2}$Large Networks and Systems Group (LANEAS), CentraleSupélec \\
		${}^{3}$Research Laboratory of Electronics, Massachusetts Institute of Technology (MIT) \\
		Email: \{marco.maso, italo.atzeni, merouane.debbah\}@huawei.com, imene.ghamnia@centralesupelec.fr, ejder@mit.edu
		\vspace{-0.2cm}
		}
	\thanks{This research has been supported by the ERC Starting Grant 305123 MORE (Advanced Mathematical Tools for Complex Network Engineering), the  U.S.  National  Science  Foundation  under  Grant CCF-1409228, and the projects 4GinVitro and BESTCOM.}
}

\begin{document}
	
	\maketitle
	
\begin{abstract}
Caching popular contents at the edge of the network can positively impact the performance and future sustainability of wireless networks in several ways, e.g., end-to-end access delay reduction and peak rate increase. In this paper, we aim at showing that non-negligible performance enhancements can be observed in terms of network interference footprint as well. To this end, we consider a full-duplex small-cell network consisting of non-cooperative cache-aided base stations, which communicate simultaneously with both downlink users and wireless backhaul nodes. We propose a novel static caching model seeking to mimic a geographical policy based on local files popularity and calculate the corresponding cache hit probability. Subsequently we study the performance of the considered network in terms of throughput gain with respect to its cache-free half-duplex counterpart. Numerical results corroborate our theoretical findings and highlight remarkable performance gains when moving from cache-free to cache-aided full-duplex small-cell networks.
\end{abstract}
	
	
\section{Introduction} \label{sec:intro}

During the last decade, the rise in smartphone usage has led to a $4000$-fold mobile data traffic increase in cellular networks \cite{Cisco2016}. This growth is expected to continue steadily in the coming years and, as a result, many operators have set the reshaping of their current mobile networks as a paramount goal. Several ambitious targets have been identified, such as higher spectral and energy efficiency, better coverage, and lower end-to-end delays: these call for significant technological advancements at both device and network level in the near future \cite{Andrews2014Will}.
	
A noteworthy example of device-level technology that recently gained momentum is the full-duplex (FD) radio \cite{Sabharwal2014InBand}. Such a device is able to  transmit and receive data over the same time/frequency resource and thus promises a non-negligible spectral and energy efficiency increase as compared to half-duplex (HD) radios \cite{Marco2015Energy}. Conversely, relevant examples of network-level solutions are ultra-dense small-cell (SC) networks. Therein, different types of SCs, i.e., micro, pico, and femto cells, are densely deployed with the objective of: \textit{(i)} reducing the distance between base stations (BSs) and user terminals (UTs) to enhance the network  coverage and spectral/energy efficiency, and \textit{(ii)} offloading the overlaying macro-cell infrastructure \cite{YunasSC}. 

Despite their individual potential, the aforementioned techniques do not offer straightforward interoperability among themselves \cite{Atz17,GoyalFDdouble}. The major issue lies in the interference footprint of FD links, which is generally much larger than its HD counterpart. In fact, the presence of several FD links potentially interfering at all times causes an overabundance of BS-to-BS and UT-to-UT interference in the network, in addition to the strong self-interference (SI) at the receive side. This heavily reduces the feasibility of the aggressive spatial frequency reuse brought by network densification. Additionally, it does not allow the theoretical throughput doubling brought by FD operations to correctly scale from device to network level, unless costly precautions are taken.
%
%
Solutions targeting a reduction of the FD interference footprint have been recently proposed to improve the FD throughput gain scalability. Most of them fall into the family of user scheduling algorithms, (self-)interference cancellation, and power control techniques, and require infrastructural changes and/or additional signaling among nodes to take place \cite{Atz17,Ale16,GoyalFDdouble,Bai2013DistributedFD,Atz16}.  

The goal of this paper is to propose an alternative to the aforementioned approaches that can improve the FD throughput gain scaling without requiring changes in terms of infrastructure or signaling. We consider an ultra-dense non-cooperative SC network where each BS is connected to the internet through a wireless backhaul (BH) node and serves one downlink UT. Similarly to \cite{Atz17,GoyalFDdouble}, we assume that BSs can operate in FD mode whereas both BHs and UTs operate in HD mode. Furthermore, we suppose that each BS has (limited) storage capabilities and can pre-fetch locally popular files before its transmit operations. System-level benefits brought by the adoption of caching have been lately studied in various settings\cite{Bas15,li2016optimization,hamidouche2016mean,yang2016interference,krishnan2016effect}. In particular, cache-aided SC networks have been shown to have a great potential to lower the BH load, in turn reducing the end-to-end access delay and increasing the peak rate \cite{Pas16}. In this context, here we aim at investigating another potential benefit brought the adoption of caching, i.e., an overall reduction of the aggregate network interference if the SC operate in FD. 

We start by designing a novel static caching policy based on local files popularity and we characterize the corresponding cache hit probability. Subsequently, we derive the probability of successfully transmitting a file in the considered random network to study the throughput gain of cache-aided FD SCs. We perform the system-level analysis using tools from stochastic geometry: this choice allows us to generalize our study and obtain tractable expressions for the performance of random FD SC network. Finally, numerical results obtained through simulations confirm the correctness of our analysis and highlight a remarkable FD throughput enhancement when moving from a cache-free to a cache-aided FD SC network.

\section{System Model} \label{sec:model}
\subsection{Network Model} \label{sec:model_network}

We consider an ultra-dense network consisting of: \textit{(i)} a set of macro-cell BHs with internet access, \textit{(ii)} a set of SC BSs ensuring network coverage, and \textit{(iii)} a set of UTs. In our model, each SC communicates with exactly one BH in the uplink (UL) and serves exactly one UT in the downlink (DL), acting as a relay between the two. The SCs have FD capabilities, whereas both BHs and UTs operate exclusively in HD mode; all communications occur in the same frequency band. Having the SCs as focus-nodes, the BHs and the UTs are thus termed as UL nodes and DL nodes, respectively.

To capture the randomness of realistic ultra-dense SC deployments and, at the same time, obtain accurate and tractable expressions for the system-level performance metrics, we model the spatial distribution of the network nodes by means of the homogeneous, independently marked Poisson point process (PPP) $\Phi \triangleq \big\{ (x, u(x), d(x))\big\} \subset \mathbb{R}^{2} \times \mathbb{R}^{2} \times \mathbb{R}^{2}$. We use $\Phi_{\sc} \triangleq \{x\}$ to denote the PPP of the SCs with spatial density $\lambda$ (measured in SCs/m$^{2}$), with isotropic marks $\Phi_{\ul} \triangleq u(\Phi_{\sc}) = \{u(x)\}_{x \in \Phi_{\sc}}$ and $\Phi_{\dl} \triangleq d(\Phi_{\sc}) = \{d(x)\}_{x \in \Phi_{\sc}}$ denoting the UL and DL nodes, respectively. Let $r_{y z} \triangleq \| y - z \|$ denote the distance between any pair of nodes $y,z \in \Phi$: the distances of the UL and DL nodes from their SCs are assumed fixed and are denoted by $R_{\ul} \triangleq r_{u(x) x}$ and $R_{\dl} \triangleq r_{x d(x)}$, $\forall x \in \Phi_{\sc} $, respectively. In this setting, it is evident that the PPPs $\Phi_{\ul}$ and $\Phi_{\dl}$ are dependent on $\Phi_{\sc}$ and are also of spatial density $\lambda$.

\subsection{From Cache-free to Cache-aided} \label{sec:model_caching}


Assume that the UL nodes have access to a \textit{global file catalog} $\mathcal{F} \triangleq \lbrace f_{1}, f_{2}, \ldots, f_{F} \rbrace$ of size $F$ files with identical lengths,\footnote{If the files have different lengths, they can be simply divided into chunks of equal length.} which represents a subset of all the files available in the internet. At a given time instant, each DL node requests a content from $\setF$: the serving SC, operating in FD mode, retrieves it from the corresponding UL node and relays it to the DL node. However, in ultra-dense network scenarios, the reliability of content transmissions may be hindered by such an aggressive time/frequency reuse: this may reduce the throughput as compared to a HD network if no proper precautions are taken.

\begin{figure}[!t]
\centering
{\def\svgwidth{\columnwidth}
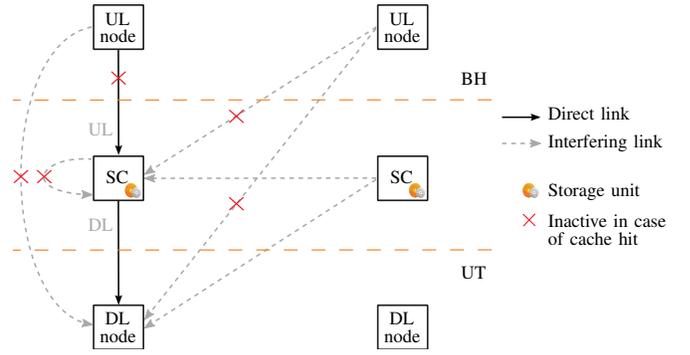}
	\caption{System model with cache-aided SCs, UL nodes, and DL nodes, with corresponding direct and interfering links.}
	\label{fig:systemmodel}
\end{figure}

Suppose now that SC $x \in \Phi_{\sc}$ is equipped with a \textit{storage unit} $\Delta_{x}$ of size $S$ files, with $S < F$, and that DL node $d(x)$ requests file $f_{i} \in \setF$. The probability of each file being requested is given by $\mathcal{P} \triangleq \lbrace p_{1}, p_{2}, \ldots, p_{F} \rbrace$, with $\sum_{i=1}^{F} p_{i} = 1$, which is determined by the files popularity over the whole network. In this context, if $f_{i}$ is cached at SC $x$, i.e., $f_{i} \in \Delta_{x}$, we have a \textit{cache hit} event and DL node $d(x)$ is served directly by SC $x$ without any backhauling transmission from UL node $u(x)$; on the contrary, if $f_{i}$ is not available, i.e., $f_{i} \not\in \Delta_{x}$, we have a \textit{cache miss} event and SC $x$ must retrieve the file from UL node $u(x)$ and relay it to $d(x)$ operating in FD mode. Hence, a cache hit event does not only allow to offload the overlaying macro-cell infrastructure, but also eliminates the need of FD operation at the SC, since the UL communication becomes inactive. The advantage in terms of interference footprint is two-fold: \textit{(i)} in a single-cell, the SI (at the SC) and the inter-node interference (at the DL node) can be eliminated; \textit{(ii)} in addition, in a multi-cell network, the interference from the other cells can be reduced. Figure~\ref{fig:systemmodel} provides a graphical representation of the resulting scenario, which is formalized in Section~\ref{sec:model_SIR}.

The effectiveness of such cache-aided approach depends on the probability that any file requested by a given DL node is cached at its serving SC, which is termed as \textit{cache hit probability} and is denoted by $\Phit$. Section~\ref{sec:P_hit} presents a novel framework to model such probability in non-cooperative random networks by capturing the local files popularity. Then, section~\ref{sec:P_suc} analyzes the system-level performance gain of cache-aided SCs in FD networks for any $\Phit$.

\subsection{Channel Model} \label{sec:model_ch}

In our model, all nodes are equipped with a single antenna for reception/transmission.\footnote{The study of the multi-antenna case goes beyond the scope of this paper and can be carried out using the framework in \cite{Atz17}.} Furthermore, the UL nodes and the SCs transmit with powers $\rho_{\ul}$ and $\rho_{\dl}$, respectively.

The propagation through the wireless channel is characterized as the combination of pathloss attenuation and small-scale fading. We introduce the pathloss function $\ell(y,z) \triangleq r_{y z}^{-\alpha}$ between nodes $y$ and $z$. In accordance with the ITU-R urban micro-cellular (UMi) pathloss model \cite{3GPP12}, which specifies different attenuations for links between different types of nodes, we fix $\alpha = \alpha_{2}$ if $y \in \Phi_{\ul} \land z \in \Phi_{\dl}$ and $\alpha = \alpha_{1}$ otherwise, where the pathloss exponents satisfy $\alpha_{2} \geq \alpha_{1} > 2$. Hence, our model admits stronger pathloss attenuation for the direct link between UL and DL nodes with respect to the other links, which is particularly suitable for our scenario where the UL nodes consist in BHs in likely non-line-of-sight with the DL nodes. Regarding the small-scale fading, we assume that all channels except the SI channel are subject to Rayleigh fading, whereas the SI channel is subject to Rician fading \cite{Dua12}. Let $h_{y z}$ denote the channel power fading gain between nodes $y$ and $z$: we have $h_{y z} \sim \chi_{2}^{2}$ if $y \neq z$ and $h_{y y} \sim \Gamma(a,b)$, where the shape and scale parameters $a$ and $b$ can be computed in closed form using \cite[Lem.~1]{Atz17} from the Rician $K$-factor and the SI attenuation measured at the SC when operating in FD mode.

\subsection{Signal-to-Interference Ratio} \label{sec:model_SIR}

In scenarios characterized by massive and dense SC deployments, such as ours, it makes sense to focus on the interference-limited regime. Hence, in the following, we formalize the signal-to-interference ratio (SIR) at the SCs and at the DL nodes. We use the notation $\slashed{\Delta}_{x}$ to denote a cache miss event at SC $x$ and we introduce the indicator function
\begin{align}
\mathbbm{1}_{\slashed{\Delta}_{x}} \triangleq \begin{cases}
1, & \mathrm{if} \ \slashed{\Delta}_{x} \\
0, & \mathrm{otherwise}.
\end{cases}
\end{align}
Hence, the SIR at SC $x$ reads as
\begin{align} \label{eq:sir_x}
\sir_{x} & \triangleq \frac{\rho_{\ul} R_{\ul}^{-\alpha_{1}} h_{u(x)x}}{I_{x}}
\end{align}
with
\begin{align} \label{eq:I_x}
I_{x} \triangleq \! \sum_{y \in \Phi_{\sc} \backslash \{x\}} \! \big( \rho_{\dl} r_{y x}^{-\alpha_{1}} h_{y x} \! + \! \rho_{\ul} r_{u(y) x}^{-\alpha_{1}} h_{u(y) x} \mathbbm{1}_{\slashed{\Delta}_{y}} \big) + h_{x x} \mathbbm{1}_{\slashed{\Delta}_{x}}.
\end{align}
Likewise, the SIR at DL node $d(x)$ is given by
\begin{align} \label{eq:sir_dx}
\sir_{d(x)} & \triangleq \frac{\rho_{\dl} R_{\dl}^{-\alpha_{1}} h_{x d(x)}}{I_{d(x)}}
\end{align}
with
\begin{align} \label{eq:I_dx}
\nonumber \hspace{-1mm} I_{d(x)} & \triangleq \! \sum_{y \in \Phi_{\sc} \backslash \{x\}} \! \big( \rho_{\ul} r_{y d(x)}^{-\alpha_{1}} h_{y d(x)} \! + \! \rho_{\dl} r_{u(y) d(x)}^{-\alpha_{1}} h_{u(y) d(x)} \mathbbm{1}_{\slashed{\Delta}_{y}} \big) \\
& \hspace{3.2cm} + \rho_{\ul} r_{u(x) d(x)}^{-\alpha_{2}} h_{u(x) d(x)} \mathbbm{1}_{\slashed{\Delta}_{x}}.
\end{align}
The beneficial effect of cache-aided SCs is clear from observing \eqref{eq:I_x} and \eqref{eq:I_dx}: in addition to reducing the aggregate network interference, a cache hit event allows to remove the SI (at the SCs) \cite{Dua12} and the inter-node interference (at the DL nodes) \cite{Ale16}, which are two main factors that hinder the practical feasibility of FD technology.

\vspace{-2mm}
\section{Caching Model} \label{sec:P_hit}

In this section, we present a non-cooperative, static caching model seeking to mimic a geographical caching policy based on local files popularity.\footnote{The study of more complex, cooperative caching policies goes beyond the scope of this paper.} This determines how files are requested by the DL nodes and cached at the SCs. We model the spatial distribution of the contents in the global file catalog $\mathcal{F}$ using the homogeneous, independently marked PPP $\Psi \triangleq \big\{ (y, f(y)) \big\} \subset \mathbb{R}^{2} \times \setF$, where $\Psi_{\setF} \triangleq \{ y \}$ denotes the PPP of the files with spatial density $\eta$ (measured in files/m$^{2}$). In this setting, each file $f_{i} \in \mathcal{F}$ yields a thinned PPP with spatial density $p_{i} \eta$. Furthermore, without loss of generality, we assume that the files in $\mathcal{F}$ are indexed by popularity, i.e., $p_{1} \geq p_{2} \geq ... \geq p_{F}$.

Let us use $\mathcal{B} (z, \varrho)$ to denote the ball of radius $\varrho$ (measured in m) centered at node $z \in \Phi_{\sc} \cup \Phi_{\dl}$. In the following, we formalize how DL nodes request files geographically and introduce the concept of \textit{request region}.

\begin{definition}[Request region] \label{def:R}
Suppose that DL node $d(x) \in \Phi_{\dl}$ is interested in requesting files that are popular in its geographical proximity. Then, the request region of DL node $d(x)$ is defined as
\begin{equation}
\mathcal{R}_{d(x)} \triangleq \big\{ \Psi_{\setF} \cap \mathcal{B}( d(x), R_{\mathrm{R}}) \big\}.
\end{equation}
\end{definition}

\begin{remark}
The radius $R_{\mathrm{R}}$ relates the local interests of the UTs to globally requested files. As $R_{\mathrm{R}} \rightarrow \infty$, and provided that $\{p_{i} > 0\}_{i=1}^{F}$, DL node $d(x)$ becomes interested in requesting all possible files in the global file catalog $\mathcal{F}$.
\end{remark}

\noindent The aim of Definition~\ref{def:R} is to capture geographical aspects of the file interests of the UTs, which are not taken into account by the existing literature (see \cite{Pas16} for an overview on content request models). Observe that, in a non-cooperative caching setting, $\Phit$ is maximized when the $S$ most popular files are cached at the SCs. Next, we formalize how SCs cache files geographically and introduce the concept of \textit{caching policy}.

\begin{definition}[Caching policy]\label{def:C}
Suppose that SC $x \in \Phi_{\sc}$ is interested in caching files that are popular in its geographical proximity. Then, a potential cache region is defined as
\begin{equation}
\mathcal{C}_{x} \triangleq \big\{ \Psi_{\setF} \cap \mathcal{B} (x, R_{\mathrm{C}}) \big\}.
\end{equation}
The caching policy of SC $x \in \Phi_{\sc}$ is defined as
\begin{equation}
\Delta_{x} \triangleq \big\{ f_i : f_i \in \mathcal{C}_{x} \land i \leq S \big\}.
\end{equation}
\end{definition}

\begin{remark}
Such caching policy allows SCs to cache only geographically close (and, therefore, popular) files, thus attempting to reduce file pre-fetching overheads in the BH.
\end{remark}

\begin{remark}
As $R_{\mathrm{C}} \rightarrow \infty$, such geographical caching policy converges to storing globally popular files as in {\rm \cite{Bas15}}.
\end{remark}

\noindent Finally, the expression of the cache hit probability under the described caching model is provided in the following Lemma.

\begin{lemma} \label{lem:P_hit}
The cache hit probability is expressed as follows:
\begin{equation} \label{eq:P_hit}
\Phit = \frac{1}{F} \sum_{i=1}^{S} \big( 1-e^{- p_{i} \eta \pi R_{\mathrm{R}}^{2}} \big) \big( 1 - e^{-p_{i} \eta \pi R_{\mathrm{C}}^{2}} \big).
\end{equation}
\end{lemma}
\begin{IEEEproof}
See Appendix~\ref{sec:app_P_hit}.
\end{IEEEproof} \vspace{1mm}

\section{Performance Analysis} \label{sec:P_suc}

In this section, we assess the system-level performance gains of the cache-aided FD SC network described in Section~\ref{sec:model}. As a performance metric, we consider the \textit{success probability}, which is defined as the probability of a DL node successfully receiving a requested file either directly from its serving SC or via the UL node. Recall that, in case of cache hit, the content transmission involves only the DL communication (i.e., from the SC to the DL node). Conversely, in case of cache miss, the content transmission occurs over the two hops (i.e., from the UL node to the DL node through the SC operating in FD mode), which creates additional interference. Our analysis focuses on a \textit{typical SC} $x$ and its marks $u(x)$ and $d(x)$, referred to as \textit{typical UL node} and \textit{typical DL node}, respectively. Due to Slivnyak's theorem and to the stationarity of $\Phi$, these nodes are representative of the whole network \cite{Hae12}.

We assume that a content is successfully transmitted over the complete communication path, i.e., from  the typical UL node to the typical DL node through the typical SC if $\sir_{x} > \theta \land \sir_{d(x)} > \theta$ for a given SIR threshold $\theta$.\footnote{Without loss of generality, we consider the same SIR threshold for both UL and DL communications.} Since the UL communication is inactive with probability $\Phit$ thanks to the caching capabilities of the typical SC, the success probability reads as
\begin{align}
\nonumber \Psuc (\theta) \triangleq \Phit & \mathbb{P}(\sir_{d(x)} > \theta) \\
\label{eq:P_suc} & + (1-\Phit) \mathbb{P}(\sir_{x} > \theta, \sir_{d(x)} > \theta).
\end{align}
Following this definition, $\Psuc (\theta)$ can be used to obtain other useful performance metrics: for instance, the achievable area spectral efficiency is readily given by $\lambda \Psuc (\theta) \log_{2} (1+\theta)$.

Let us provide the following preliminary definitions: 
\begin{align}
\label{eq:upsilon_hat} \widehat{\Upsilon} (s) & \triangleq \frac{\pi (s \rho_{\dl})^{\frac{2}{\alpha_{1}}} \csc \big( \frac{2 \pi}{\alpha_{1}} \big)}{\alpha_{1}} \\
\widetilde{\Upsilon} (s) & \triangleq \int_{0}^{\infty} \bigg( 1 - \frac{1}{1 + s \rho_{\dl} r^{- \alpha_{1}}} \Omega (s,r) \bigg) r \diff r \\
\Omega (s,r) & \triangleq \frac{1}{2 \pi} \int_{0}^{2 \pi} \frac{\diff \varphi}{1 + s \rho_{\ul} (R_{\ul}^{2} + r^{2} + 2 R_{\ul} r \cos \varphi)^{- \frac{\alpha_{2}}{2}}}
\end{align}
Recall the expressions of $I_{x}$ and $I_{d(x)}$ in \eqref{eq:I_x} and \eqref{eq:I_dx}, respectively. Next theorem provides a tight lower bound on the $\Psuc (\theta)$ in \eqref{eq:P_suc}, with some properties given in Corollary~\ref{cor:P_suc}.

\begin{theorem} \label{th:P_sucLB}
The success probability in \eqref{eq:P_suc} is bounded as $\Psuc (\theta) \geq \PsucLB (\theta)$, with
\begin{align}
\nonumber \PsucLB (\theta) & \triangleq \Phit \setL_{I_{d(x)}} (\theta \rho_{\dl}^{-1} R_{\dl}^{\alpha_{1}}) \\
\label{eq:P_sucLB} & + (1 - \Phit) \setL_{I_{x}}^{\slashed{\Delta}_{x}} (\theta \rho_{\ul}^{-1} R_{\ul}^{\alpha_{1}}) \setL_{I_{d(x)}}^{\slashed{\Delta}_{x}} (\theta \rho_{\dl}^{-1} R_{\dl}^{\alpha_{1}})
\end{align}
where $\setL_{I_{d(x)}} (s)$ is the Laplace transform of the interference at DL node $d(x)$ in case of cache hit, whereas $\setL_{I_{x}}^{\slashed{\Delta}_{x}} (s)$ and $\setL_{I_{d(x)}}^{\slashed{\Delta}_{x}} (s)$ are the Laplace transforms of the interference at SC $x$ and at DL node $d(x)$, respectively, in case of cache miss:
\begin{align}
\nonumber \mathcal{L}_{I_{d(x)}} (s) & \triangleq \exp \big( - 2 \pi \lambda \Phit \widehat{\Upsilon} (s) \big) \\
\label{eq:L_dx} & \phantom{=} \ \times \exp \big( - 2 \pi \lambda (1 - \Phit) \widetilde{\Upsilon} (s) \big) \\
\label{eq:L_x_D} \mathcal{L}_{I_{x}}^{\slashed{\Delta}_{x}} (s) & \triangleq \frac{1}{(1 + s \rho_{\dl} b)^{a}} \mathcal{L}_{I_{d(x)}}(s) \\
\label{eq:L_dx_D} \mathcal{L}_{I_{d(x)}}^{\slashed{\Delta}_{x}} (s) & \triangleq \Omega (s, R_{\dl}) \setL_{I_{d(x)}}(s).
\end{align}
\end{theorem}

\begin{IEEEproof}
See Appendix~\ref{sec:app_P_suc}.
\end{IEEEproof} \vspace{2mm}

\begin{corollary} \label{cor:P_suc}
The lower bound on the success probability in \eqref{eq:P_sucLB} enjoys the following properties:
\begin{itemize}
\item[(a)] $\PsucLB (\theta) \to \Psuc (\theta)$ as $\Phit \to 1$;
\item[(b)] $\PsucLB (\theta) = \Psuc (\theta)$ in case of uncorrelated locations of the nodes between UL and DL communications.
\end{itemize}
\end{corollary}

\begin{IEEEproof}
The proof is straightforward from Appendix~\ref{sec:app_P_suc}.
\end{IEEEproof} \vspace{2mm}

\bgroup
\def\arraystretch{1.6}%
\begin{table}[!t]
	\caption{Simulation Parameters.}
	\label{tab:simparams}
	\centering
	\small
	\scalebox{0.8}{
		\begin{tabular}{|c|c|c|}
			\hline
			{\bf System Parameter}		& {\bf Symbol} 			& {\bf Value}\\
			\hline
			\hline
			\hline
			Storage to Catalog size ratio 			 	& $\kappa$ 					& $\{0.1, 0.35, 0.6\}$  \\
			Catalog shape parameter 	& $\gamma$ 				& $0.7$ 				\\
			Radius of request region 	& $R_{\mathrm{R}}$ 			& $8$ $\mathrm{m}$ \\
			Radius of cache region 	& $R_{\mathrm{C}}$			& $40$ $\mathrm{m}$ \\
			\hline
			\hline
			Distance of UL node 	& $R_{\mathrm{UL}}$  	& $20$  $\mathrm{m}$ \\
			Distance of DL node 	& $R_{\mathrm{DL}}$  	& $5$ $\mathrm{m}$ \\
			Transmit power of UL node   & $\rho_{\mathrm{UL}}$  & $1$ $\mathrm{W}$ \\
			Transmit power of DL node   & $\rho_{\mathrm{DL}}$ & $0.2$ $\mathrm{W}$ \\
			\hline
			\hline
			Path-loss exponent I 		& $\alpha_1$ 		& $3$			 \\
			Path-loss exponent II		& $\alpha_2$			& $4$				 \\
			Rician $K$-factor 			& $K$ 					& $1$ 				  \\
			Self-interference attenuation &  					& $80$ $\mathrm{dB}$ \\
			Target SIR   				& $\theta$ 				& $0$ $\mathrm{dB}$ \\
			\hline
		\end{tabular}
	}
\end{table}
\bgroup
\def\arraystretch{1.0}%

\section{Numerical Results} \label{sec:num}

The analytical expressions derived in Sections~\ref{sec:P_hit} and \ref{sec:P_suc} are validated here by means of Monte Carlo simulations. 
As conventionally done in the literature (cf. \cite{Pas16}), we assume that the global file catalog follows a Zipf popularity distribution, i.e., each file $f_{i} \in \mathcal{F}$ has popularity $p_{i} \in \mathcal{P}$ defined as
\begin{equation}
p_{i} =  \bigg ( i^{\gamma} \sum_{j=1}^{F} \frac{1}{j^{\gamma}} \bigg )^{-1}
\end{equation}
where $\gamma$ is the catalog shape parameter. The contents in the global file catalog are then proactively cached by the SCs based on the caching policy defined in Definition~\ref{def:C} and the UTs are served accordingly. For notational simplicity, let $\kappa \triangleq \frac{S}{F} \leq 1$ denote the ratio between storage size and catalog size. The set of simulation parameters is listed in Table~\ref{tab:simparams} and is considered unless stated otherwise; observe that the parameters $a$ and $b$ in \eqref{eq:L_x_D}, related to the SI power at the FD SCs, are computed as in \cite[Lem.~1]{Atz17} using the Rician $K$- factor and the SI attenuation. Furthermore, let us introduce the \textit{FD throughput gain} defined as
\begin{align}
\label{eq:TG} \mathsf{TG}_{\mathrm{FD}} (\theta) \! \triangleq \! 2 \Psuc (\theta) \exp \bigg( \! 2 \pi \lambda \frac{\pi \theta^{\frac{2}{\alpha_{1}}} \big( R_{\ul}^{2} + R_{\dl}^{2} \big) \csc \big( \frac{2 \pi}{\alpha_{1}} \big)}{\alpha_{1}} \! \bigg)
\end{align}
which quantifies the throughput gain of cache-aided SC networks in FD mode with respect to cache-free SC networks in HD mode. This performance metric relates the success probability $\Psuc (\theta)$ in \eqref{eq:P_suc} with the success probability of an equivalent cache-free HD setting \cite{Atz17} taking into account the double spectral efficiency gain brought by FD mode: in particular $\mathsf{TG}_{\mathrm{FD}} (\theta) > 1$ indicates that FD outperforms the equivalent HD setup.

\begin{figure}
	\includegraphics[width=\columnwidth]{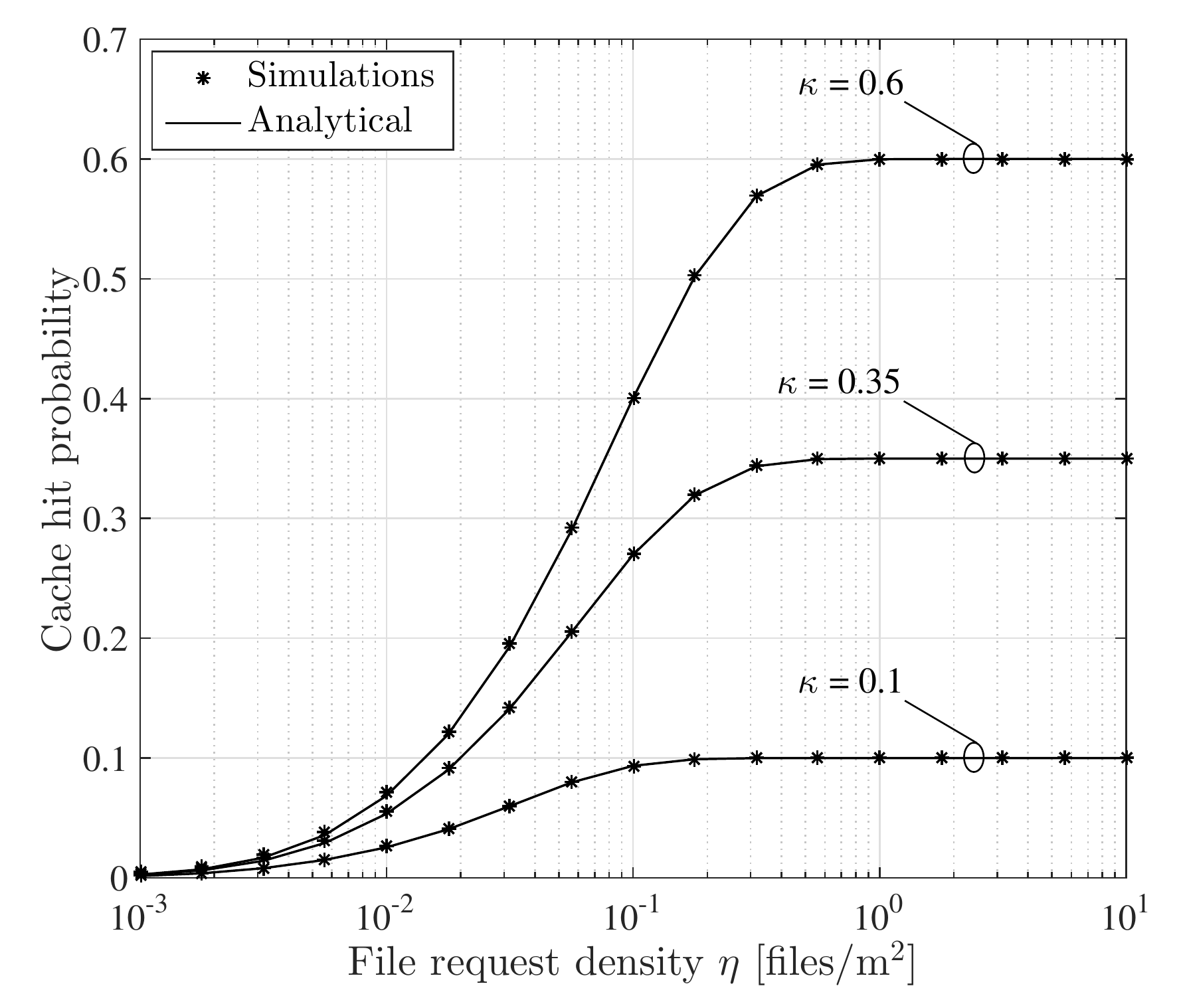}
	\caption{The impact of file request density $\eta$ on the cache hit probability.} \vspace{-2mm}
	\label{fig:phitVSfilerequest}
\end{figure}

The cache hit probability $\mathsf{P}_{\mathrm{hit}}$ in \eqref{eq:P_hit} is illustrated in Figure~\ref{fig:phitVSfilerequest} as a function of the file request density $\eta$. On the one hand, the analytical expression in Lemma~\ref{lem:P_hit} accurately matches the obtained numerical results. On the other hand, for any given $\kappa$, increasing the file request density yields a higher $\mathsf{P}_{\mathrm{hit}}$. In fact, a higher file request density allows to capture the files popularity more accurately, in turn improving the efficiency of the storage use. In other words, as $\eta$ increases, the cached files statistics converge to the file requests statistics. Furthermore, the impact of non-sufficient sampling of the files popularity is especially evident for high values of $\kappa$ (i.e., when the SCs are equipped with large storage units).

Recall that, based on Corollary~\ref{cor:P_suc}, $\PsucLB (\theta)$ in \eqref{eq:P_sucLB} gives the exact expression of $\Psuc (\theta)$ in \eqref{eq:P_suc} in case of uncorrelated locations of the nodes between UL and DL communications. The FD throughout gain \eqref{eq:TG} and its lower bound obtained using $\PsucLB (\theta)$ are depicted in Figure~\ref{fig:psuccessVSnodedensity} as functions of the SC density $\lambda$ for a fixed file request density $\eta=1$~files/m$^{2}$: in this regard, the curves for $\kappa=0$ corresponds to the cache-free case studied in \cite{Atz17}. On the one hand, the analytical lower bound is remarkably tight and it is increasingly accurate as $\kappa$ grows (this is in accordance with Corollary~\ref{cor:P_suc}). On the other hand, the throughput gain steadily increases with $\kappa$ for any value of the SC density $\lambda$: this stems from the fact that storing popular contents at the SCs yields a considerable reduction in the aggregate network interference as compared to the cache-free case. For instance, for $\lambda=10^{-4}$~SCs/m$^{2}$, we have $\mathsf{TG}_{\mathrm{FD}} (\theta) = 1.7$ with $\kappa=0$ and $\mathsf{TG}_{\mathrm{FD}} (\theta) = 1.85$ with $\kappa=0.6$; moreover, for $\lambda=10^{-3}$~SCs/m$^{2}$, we have $\mathsf{TG}_{\mathrm{FD}} (\theta) = 0.42$ with $\kappa=0$ and $\mathsf{TG}_{\mathrm{FD}} (\theta) = 1.11$ with $\kappa=0.6$.
Therefore, adding adequate caching capabilities at the SCs allows to improve the area spectral efficiency by operating the network in FD mode at higher SC densities. This gives a very clear insight to the network planners for selecting appropriate deployment strategies: in this respect, the tradeoff is given by SC density $\lambda$ against storage size $S$ depending on the deployment cost of each aspect. 


\begin{figure}[!t]
	\includegraphics[width=\columnwidth]{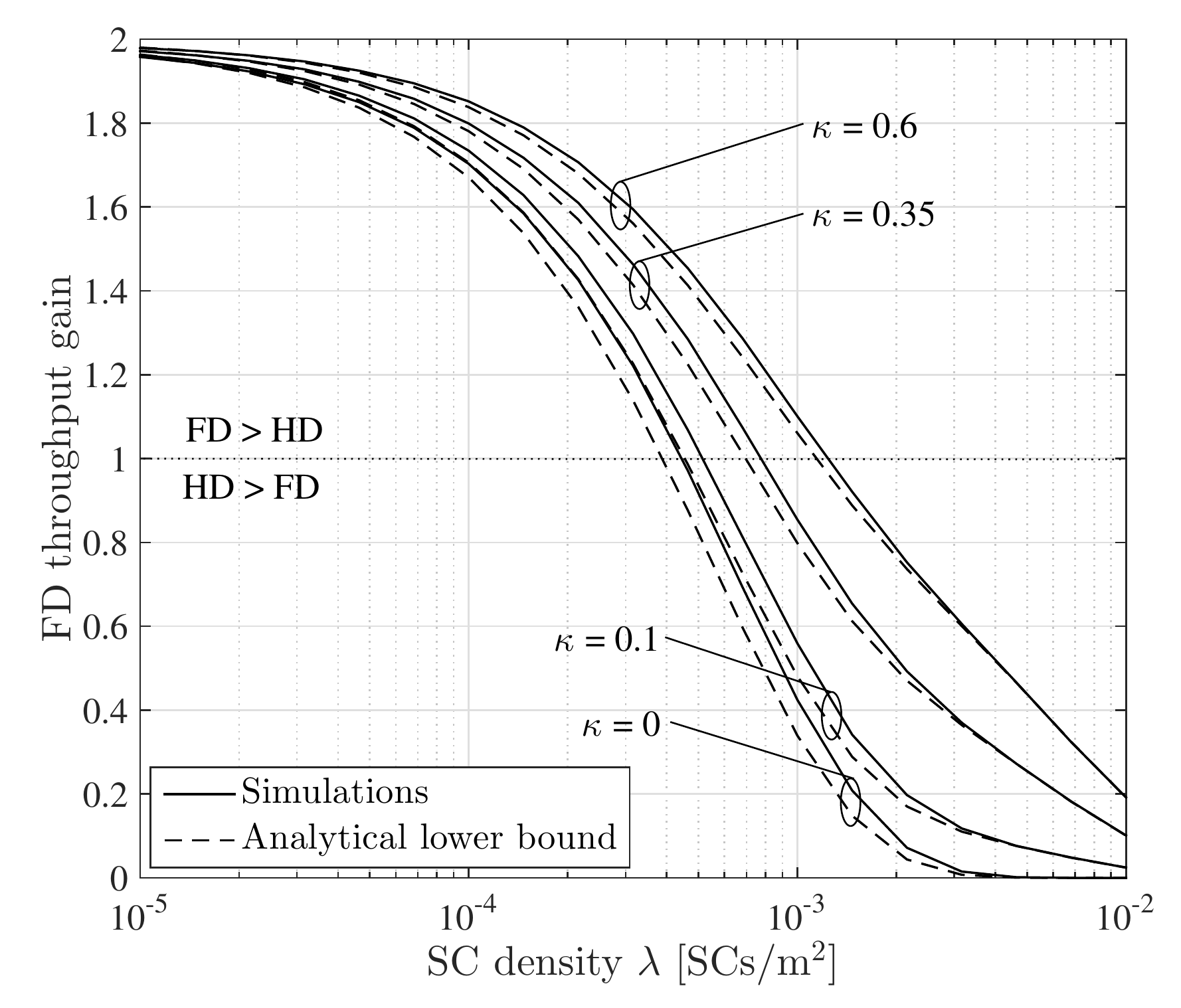}
	\caption{The impact of SC density $\lambda$ on the FD throughput gain.} \vspace{-2mm}
	\label{fig:psuccessVSnodedensity}
\end{figure}


\begin{figure*}
	\begin{align} \addtocounter{equation}{+1}
	\label{eq:L_dx1} \setL_{I_{d(x)}} (s) & = \Exp [e^{- s I_{d(x)}}] \\
	& = \Exp \bigg[ \exp \bigg( - s \sum_{y \in \Phi_{\sc} \backslash \{x\}} \big( \rho_{\dl} r_{y d(x)}^{- \alpha_{1}} h_{y d(x)} + \rho_{\ul} r_{u(y) d(x)}^{- \alpha_{1}} h_{u(y) d(x)} \mathbbm{1}_{\slashed{\Delta}_{y}} \big) \bigg) \bigg] \\
	& = \Exp \bigg[ \prod_{y \in \Phi_{\sc} \backslash \{x\}} \exp \bigg( - s \big( \rho_{\dl} r_{y d(x)}^{- \alpha_{1}} h_{y d(x)} + \rho_{\ul} r_{u(y) d(x)}^{- \alpha_{1}} h_{u(y) d(x)} \mathbbm{1}_{\slashed{\Delta}_{x}} \big) \bigg) \bigg] \\
	& = \Exp \bigg[ \prod_{y \in \widehat{\Phi}_{\sc} \backslash \{x\}} \! \exp \big( - s \rho_{\dl} r_{y d(x)}^{- \alpha_{1}} h_{y d(x)} \big) \bigg] \Exp \bigg[ \prod_{y \in \widetilde{\Phi}_{\sc} \backslash \{x\}} \! \exp \bigg( - s \big( \rho_{\dl} r_{y d(x)}^{- \alpha_{1}} h_{y d(x)} + \rho_{\ul} r_{u(y) d(x)}^{- \alpha_{1}} h_{u(y) d(x)} \big) \bigg) \bigg] \\
	\label{eq:L_dx2} & = \exp \bigg( - 2 \pi \lambda \Phit \int_{0}^{\infty} \bigg( 1 - \frac{1}{1 + s \rho_{\dl} r^{- \alpha_{1}}} \bigg) r \diff r \bigg) \exp \bigg( - 2 \pi \lambda (1 - \Phit) \int_{0}^{\infty} \bigg( 1 - \frac{1}{1 + s \rho_{\dl} r^{- \alpha_{1}}} \Omega (s,r) \bigg) r \diff r \bigg)
	\end{align} \addtocounter{equation}{-6}
	\hrulefill \vspace{-3mm}
\end{figure*}

\section{Conclusions} \label{sec:concl}

This paper considers an ultra-dense network consisting of small cells (SCs) with full-duplex (FD) and caching capabilities communicating simultaneously with both downlink users and wireless backhaul nodes. First, we propose a novel static caching model seeking to mimic a geographical caching policy based on local files popularity and calculate the corresponding cache hit probability. Then, using tools from stochastic geometry, we study the system-level impact of caching on FD SC networks. Remarkably, the employment of cache-aided SCs in FD scenarios allows to reduce the aggregate network interference, with a significant increase in the network performance. In particular, moving from cache-free to cache-aided SC networks allows to improve the area spectral efficiency by operating the network in FD mode at higher SC densities.

\appendices

\section{Proof of Lemma~\ref{lem:P_hit}} \label{sec:app_P_hit}

Since each file $f_i$ is distributed according to a thinned PPP with density $p_i \eta$, the probabilities of $f_i$ falling independently into the request region $\mathcal{R}_{d(x)}$ and the potential cache region $\mathcal{C}_{x}$ are readily given by $1-e^{- p_{i} \eta \pi R_{\mathrm{R}}^{2}}$ and $1 - e^{-p_{i} \eta \pi R_{\mathrm{C}}^{2}}$, respectively. Now, assuming unlimited storage at the FD SCs, the cache hit probability of file $f_i$ is calculated as the probability of file $f_i$ falling into both request and potential cache regions, which is given by $\big(1-e^{- p_{i} \eta \pi R_{\mathrm{R}}^{2}} \big) \big(1 - e^{-p_{i} \eta \pi R_{\mathrm{C}}^{2}} \big)$. Finally, considering all contents in the global file catalog $\mathcal{F}$ and imposing storage constrains from Definition~\ref{def:C} yields the cache hit probability in \eqref{eq:P_hit}.

\section{Proof of Theorem~\ref{th:P_sucLB}} \label{sec:app_P_suc}

The analytical expression \eqref{eq:P_sucLB} is constructed by assuming uncorrelated locations of the UL and DL nodes when a cache miss event occurs. Building on the Fortuin-Kasteleyn-Ginibre (FKG) inequality, such uncorrelated case represents a lower bound on the network performance for the correlated case \cite{Atz17,Hae12}. Hence, given $\Psuc (\theta)$ in \eqref{eq:P_suc}, we can write
\begin{align} \label{eq:P_suc1}
\hspace{-1mm} \Psuc (\theta) \geq \Phit \mathsf{P}_{\mathrm{suc},2}(\theta) + (1-\Phit) \mathsf{P}_{\mathrm{suc},1}^{\slashed{\Delta}_{x}}(\theta) \mathsf{P}_{\mathrm{suc},2}^{\slashed{\Delta}_{x}}(\theta)
\end{align}
where $\mathsf{P}_{\mathrm{suc},2}(\theta)$ is the probability of successfully transmitting a content from the typical SC to the typical DL node in case of cache hit, whereas $\mathsf{P}_{\mathrm{suc},1}^{\slashed{\Delta}_{x}}(\theta)$ (resp. $\mathsf{P}_{\mathrm{suc},2}^{\slashed{\Delta}_{x}}(\theta)$) is the probability of successfully transmitting a content from the typical UL node to the typical SC (resp. from the typical SC to the typical DL node) in case of cache miss.

We begin by focusing on $\mathsf{P}_{\mathrm{suc},2}(\theta)$, which can be obtained as the Laplace transform of the interference $I_{d(x)}$ in \eqref{eq:I_dx} in absence of inter-node interference, denoted by $\setL_{I_{d(x)}} (s)$. Let $\widetilde{\Phi}_{\sc} \triangleq \{ x \in \Phi_{\sc} : \slashed{\Delta}_{x} \}$ and let $\widehat{\Phi}_{\sc} \triangleq \Phi_{\sc} \backslash \widetilde{\Phi}_{\sc}$: by definition, $\widehat{\Phi}_{\sc}$ and $\widetilde{\Phi}_{\sc}$ are independent PPPs with densities $\Phit \lambda$ and $(1-\Phit) \lambda$, respectively. Hence, $\setL_{I_{d(x)}} (s)$ in \eqref{eq:L_dx} is derived as in \eqref{eq:L_dx1}--\eqref{eq:L_dx2} at the top of the page, where the integral in the first exponential has a closed-form solution given by \eqref{eq:upsilon_hat}.
On the other hand, $\mathsf{P}_{\mathrm{suc},1}^{\slashed{\Delta}_{x}}(\theta)$ (resp. $\mathsf{P}_{\mathrm{suc},2}^{\slashed{\Delta}_{x}}(\theta)$) can be obtained as the Laplace transform of the interference $I_{x}$ in \eqref{eq:I_x} (resp. $I_{d(x)}$ in \eqref{eq:I_dx}) in presence of SI \cite[Th.~1]{Atz17} (resp. inter-node interference \cite[Th.~2]{Atz17}), which is given by $\setL_{I_{x}}^{\slashed{\Delta}_{x}}(s)$ in \eqref{eq:L_x_D} (resp. $\setL_{I_{d(x)}}^{\slashed{\Delta}_{x}}(s)$ in \eqref{eq:L_dx_D}). This concludes the proof. \hfill \IEEEQED

\addcontentsline{toc}{chapter}{References}
\bibliographystyle{IEEEtran}
\bibliography{IEEEabrv,references,references-caching}

\end{document}